\documentstyle[aps,prb,psfig,multicol]{revtex}
\begin{document}
\draft\flushcolumns
  \title{Nearly universal crossing point of the specific heat curves
    of Hubbard models}
  \author{N.~Chandra,\cite{Chandra98} M.~Kollar,
  D.~Vollhardt}
  \address{Theoretische Physik III, Elektronische Korrelationen und
    Magnetismus, Institut f\"{u}r Physik, Universit\"{a}t Augsburg,
    86135~Augsburg, Germany}
  \date{October 28, 1998}
  \maketitle  
  \begin{abstract} A nearly universal feature of the specific
    heat curves $C(T,U)$ vs.\ $T$ for different $U$ of a general class
    of Hubbard models is observed. That is, the value $C_+$ of the
    specific heat curves at their high-tem\-per\-a\-ture crossing
    point $T_+$ is almost independent of lattice structure and spatial
    dimension $d$, with $C_+/k_B\approx0.34$. This surprising feature
    is explained within second order perturbation theory in $U$ by
    identifying two small parameters controlling the value of $C_+$:
    the integral over the deviation of the density of states
    $N(\epsilon)$ from a constant value, characterized by $\delta
    N=\int\!d\epsilon\,|N(\epsilon)-\frac{1}{2}|$, and the inverse
    dimension, $1/d$.
  \end{abstract}

  \begin{multicols}{2}
  \section{Introduction}
  \label{intro}

\begin{figure}
    \centerline{\psfig{file=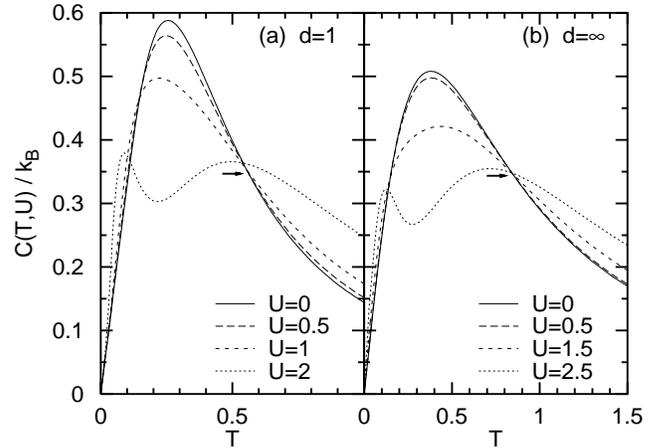,width=1.05\hsize,silent=}}%
    \caption{\global\columnwidth20.5pc
      Specific heat for the Hubbard model.  
      (a) Exact solution for $1/r$ hopping in
      $d=1$.\protect\cite{Gebhard91} (b) Iterated perturbation theory
      for NN hopping in $d=\infty$.\protect\cite{Georges93} In (a) $T$
      and $U$ are in units of the half bandwidth, while for (b) the
      second moment of the (Gaussian) density of states is set to
      unity. At the high-tem\-per\-a\-ture crossing point the specific
      heat has the almost universal value of $0.34k_B$ in the limit
      $U\to0$ (see arrows).\label{cpic}}
  \end{figure}

  Recently attention was drawn to the fact that in various strongly
  correlated systems the curves of the specific heat $C(T,X)$ vs.\ 
  temperature $T$ cross once or twice when plotted for different
  values of a second thermodynamic variable $X$.\cite{Vollhardt97} For
  example, crossing points are observed for different pressures
  ($X=P$) in normalfluid $^3$He (Ref.~\onlinecite{Brewer59}) and
  heavy-fermion systems such as CeAl$_3$\cite{Brodale86} and
  UBe$_3$.\cite{Phillips86} By changing the magnetic field ($X=B$) the
  same feature is seen in heavy-fermion compounds such as
  CeCu$_{6-x}$Al$_x$\cite{Schlager93} and
  Nd$_{2-x}$Ce$_x$CuO$_4$.\cite{Brugger93} Crossings of the specific
  heat curves are also observed in the simplest lattice model for
  correlated electrons, the Hubbard model,\cite{Hubbard63}
  \begin{equation}
    \hat{H} = \sum_{{\bf k}\sigma}
    (\epsilon_{\bf k}-\mu)\,
    \hat{a}^+_{{\bf k}\sigma}
    \hat{a}^{\phantom{+}}_{{\bf k}\sigma}
    + U \sum_i
    \hat{n}^{\phantom{+}}_{i\uparrow}
    \hat{n}^{\phantom{+}}_{i\downarrow},
  \end{equation}
  where $\epsilon_{\bf k}$ is the dispersion of a single electronic
  band, $\mu$ the chemical potential, and $U$ the local interaction.
  At half filling the curves $C(T,U)$ vs. $T$ always cross at two
  temperatures. This is observed, for example, in the case of the
  model with nearest-neighbor hopping in $d=1$,
  \cite{Shiba72,Juettner98} $d=2$,\cite{Duffy97} and
  $d=\infty$,\cite{Georges93} as well as for long-range hopping in
  $d=1$;\cite{Gebhard91} for the latter two systems the specific heat
  is shown in Fig.\ \ref{cpic}. Furthermore, crossing is found in
  $d=1$ when a magnetic field $B$ is changed at constant
  $U$.\cite{Usuki90} The fact that these crossing points may be very
  sharp was analyzed in Ref.~\onlinecite{Vollhardt97}, and was traced
  to the properties of certain generalized susceptibilities of the
  system.
  
  In the following, we will consider only the crossing of the specific
  heat curves occuring for $X=U$ in the paramagnetic phase of the
  Hubbard model with a symmetric half-filled band ($n=1$).  We will
  investigate yet another observation, namely that for small $U$ the
  specific heat at the high-tem\-per\-a\-ture crossing point {\em has
    practically the same value of approximately $0.34k_B$ for all
    dimensions $d$ and dispersions $\epsilon_{\bf k}$}, which can be
  seen also in Fig.\ \ref{cpic}.  This is surprising, because the
  temperatures at which this crossing occurs are very different for
  different dispersions and dimensions, and because the maximum value
  of $C(T,U)$ and its value at the low-tem\-per\-a\-ture crossing
  point vary strongly, too. It should be noted that the specific heat,
  like the entropy, is a dimensionless quantity when expressed in
  units of $k_B$.
  
  We denote by $T_+$ the temperature at which the curves $C(T,U)$ vs.\ 
  $T$ cross for different values of $U$.  Then the specific heat is
  {\em independent} of $U$ at the crossing temperature $T_+(U)$,
  defined by\cite{Vollhardt97}
  \begin{equation}
    \left.\frac{\partial C}{\partial U}\right|_{T_+(U)}=0
    \label{txudef}
  \end{equation}
  Since we are not concerned with the dependence of $T_+(U)$ on $U$,
  but rather with the crossing point value of $C(T,U)$ for different
  lattice systems, we consider only the limit of small $U$ and define
  \begin{eqnarray}
    T_+&\equiv&\lim_{U\to0^+}  T_+(U),
    \;\label{txdef}\\
    C_+&\equiv&\lim_{U\to0^+}C(T_+(U),U).
    \;\label{cxdef}
  \end{eqnarray}
  This weak-coupling crossing point can be calculated without
  approximation using second order perturbation theory in $U$. Higher
  orders in perturbation theory would be necessary to determine the
  dependence of $C(T_+(U),U)$ on $U$.
  
  It is the purpose of this paper to show that the value of $C_+$ at
  the high temperature crossing point is almost universal, and to
  analyze the origin of this peculiar feature.  We calculate $C_+$ for
  a half-filled band with a symmetric density of states (DOS)
  $N(\epsilon)=\int\!d{\bf k}\,\delta(\epsilon-\epsilon_{{\bf k}})$.
  We also show that the weak dependence of $C_+$ on lattice properties
  can be understood by starting from the limit $d=\infty$ and using
  expansions in terms of two small parameters.

  This paper is organized as follows.  In Sec.~\ref{xings} we review
  the origin of crossing points in the Hubbard model and show how to
  calculate $T_+$ and $C_+$ in second order perturbation theory, the
  details of which are contained in Appendix~\ref{e2appendix}.
  Various non-in\-ter\-act\-ing lattice systems are listed in
  Sec.~\ref{nonintsystems}, and the values of $C_+$ at the
  high-tem\-per\-a\-ture crossing point for these systems are
  presented in Sec.~\ref{cxresults}.  Sec.~\ref{theory} contains
  expansions of $C_+$ that reveal the influence of the density of
  states and the lattice dimension.  We close with a conclusion in
  Sec.~\ref{conclusion}.

  \section{Crossing points in the specific heat of the Hubbard model}
  \label{xings}

  The entropy per lattice site $S(T,U)$ is given by \begin{equation}
  S(T, U) = \int\limits_0^T{\!}dT'\;\frac{C(T', U)}{T'} .
  \;\label{entropy} \end{equation} For the Hubbard model $S(T,U)$
  approaches a constant when $T\rightarrow\infty$.  Taking the
  derivative of Eq.\ (\ref{entropy}) with respect to $U$ we find
  \begin{equation} 0 = \int\limits_0^{\infty}{\!}\frac{dT}{T}
  \;\frac{\partial C(T,U)}{\partial U}.  \;\label{sum-rule}
  \end{equation} Since $\partial C/\partial U$ is not identically
  zero, there must exist temperature regions where it has positive and
  negative values. We assume that no phase transitions occur, so that
  $\partial C/\partial U$ is a continuous function of $U$.  Then there
  exist temperatures where $\partial C/\partial U$ changes sign; at
  these temperatures the curves $C(T,U)$ vs.\ $T$ cross [see
  Eq.\ (\ref{txudef})].
  
  There are two such crossing points in the paramagnetic phase of the
  half-filled Hubbard model, as can be seen from the sign of $\partial
  C/\partial U$ at very low and very high temperatures. For
  intermediate values of $U$ the specific heat of this model shows the
  following general features: $C(T,U)$ starts linear in $T$ at low
  temperatures and develops a two-peak structure, one at temperatures
  $T \sim 4t^2/U$ due to spin excitations ($t$ is the hopping
  amplitude), and one at temperatures $T \sim U-W$ due to charge
  excitations, where $W$ is the bandwidth.  For low temperatures the
  spin exitations become stronger for increasing $U$, thus $\partial
  C/\partial U>0$.  At high temperatures an increase in $U$ pushes out
  the charge peak and thus increases $C(T,U)$, which tends to
  $(a+b\,U^2+O(U^4))/T^2$ with $a,b>0$.  Hence $\partial C/\partial
  U>0$ for both high and low temperatures, so that the sum rule [Eq.\ 
  (\ref{sum-rule})] yields an intermediate region with $\partial
  C/\partial U<0$. There are thus two sign changes corresponding to
  two crossing points.
  
  To determine the location of the crossing points defined by
  Eqs.\ (\ref{txdef}) and (\ref{cxdef}), we calculate the internal
  energy per lattice site in perturbation theory in $U$,
  \begin{equation}
    E(T,U)=
    E^{(0)}(T)
    +\frac{1}{4}\,U 
    +U^2\,E^{(2)}(T)
    +{{O}}(U^3).
    \;\label{e-expansion}
  \end{equation}
  Here ($k_B\equiv1$, $\beta=1/T$)
  \begin{equation}
    E^{(0)}(T)=2\int{\!\!}d\epsilon\;\frac{N(\epsilon)\epsilon}
    {1+\exp(\beta\epsilon)},
    \;\label{e0}
  \end{equation}
  which is the internal energy for the non-in\-ter\-act\-ing system, $U/4$
  is the Hartree contribution, and the second-order correlation energy
  is given by (see Appendix \ref{e2appendix} for details)
  \end{multicols}
  \vspace*{-3mm}\noindent\rule[2mm]{86.36mm}{.1mm}\rule[2mm]{.1mm}{2mm}\vspace*{-2mm}
  \begin{equation}
    E^{(2)}(T)=
      -\frac{{\partial}}{{\partial}\beta}\frac{\beta^2}{32} 
      \int\limits_0^{\mbox{\ }1}{\!\!}dx
      \int{\!\!}d{\bf k}
      \int{\!\!}d{\bf p}
      \int{\!\!}d{\bf q}
    \frac{\cosh[\frac{1}{2}x\beta
      (\epsilon_{\bf k} 
      +\epsilon_{\bf p}
      +\epsilon_{{\bf p}+{\bf q}}
      +\epsilon_{{\bf k}+{\bf q}})]}{
      \cosh(\frac{1}{2}\beta\epsilon_{\bf k})
      \cosh(\frac{1}{2}\beta\epsilon_{\bf p})
      \cosh(\frac{1}{2}\beta\epsilon_{{\bf k}+{\bf q}})
      \cosh(\frac{1}{2}\beta\epsilon_{{\bf p}+{\bf q}})}
    ,\!\!\!\label{e2-finite-d}
  \end{equation}
  \noindent\hspace{22pc}\vspace*{-2mm}\rule[0mm]{.1mm}{2mm}\rule[2mm]{86.36mm}{.1mm}\vspace*{-3mm}
  \begin{multicols}{2}
  where the integrations, e.~g.\ $\int{\!}  d{\bf k}\,
  \equiv\int{\!}{d^{\,d}k}/{(2\pi)^d}$, run over the first Brillouin
  zone.
  
  In the limit of infinite spatial dimensions\cite{Metzner89} this
  expression can be simplified further. In this case, momentum
  conservation at vertices becomes
  irrelevant,\cite{Mueller-Hartmann89} so that the integrals factorize
  (see Appendix \ref{e2appendix})
  \begin{eqnarray}
    E^{(2)}(T)&=& 
    -\frac{{\partial}}{{\partial}\beta}\frac{\beta^2}{32}
    \int\limits_0^{\mbox{\ }1}{\!\!}dx
    \left[\int{\!\!}d\epsilon\,N(\epsilon) 
      \frac{\cosh(\frac{1}{2}x\beta\epsilon)}{
        \cosh(\frac{1}{2}\beta\epsilon)}\right]^4
    .
    \label{e2-infinite-d}
  \end{eqnarray}
  Note that as usual in infinite dimensions the dispersion
  $\epsilon_{\bf k}$ enters into one-particle quantities only via the
  DOS $N(\epsilon)$.  Therefore this expression is much easier to
  evaluate numerically than Eq.\ (\ref{e2-finite-d}).
  
  The specific heat $C(T,U)=\partial E/\partial T$ has the expansion
  \begin{equation}
    C(T,U)=
    C^{(0)}(T)
    +U^2\,C^{(2)}(T)
    +{{O}}(U^4)
    ,
    \label{c-expansion}
  \end{equation}
  where
  \begin{eqnarray}
    C^{(0)}(T)&=&
    \frac{\beta^2}{2}
    \int{\!}d\epsilon\,
    \frac{N(\epsilon)\epsilon^2}{
      \cosh^2(\frac{1}{2}\beta\epsilon)},
    \label{c0-T}
  \end{eqnarray} 
  and the function $C^{(2)}(T)$ can be written as (see Appendix
  \ref{e2appendix})
  \begin{eqnarray}
    C^{(2)}(T)&=&
    \frac{\beta^2}{32}\frac{\partial^2}{\partial\beta^2}\;\beta^2
    \int\limits_0^{\mbox{\ }1}{\!\!}dx \sum_m \left[f_m(x,\beta)\right]^4
    .
    \label{c2-T}
  \end{eqnarray}
  Here the sum runs over lattice sites ${\bf R}_m$ and the functions
  $f_m(x,\beta)$ are given by
  \begin{eqnarray} 
    f_m(x,\beta)&=&\int{\!\!}d{\bf k}\,
    \frac{\exp\left(i\,{\bf k}\!\cdot\!{\bf R}_m
        +\frac{1}{2}x\beta\epsilon_{{\bf k}}\right)}{
      \cosh(\frac{1}{2}\beta\epsilon_{{\bf k}})}
    .
    \label{fdef}
  \end{eqnarray}
  Comparison with Eq.\ (\ref{e2-infinite-d}) shows that in $d=\infty$
  only the local term with ${\bf R}_m=0$, i.~e.\ $f_0(x,\beta)$ $=$
  $\int{\!}d\epsilon$ $N(\epsilon)$ $\cosh(\frac{1}{2}x\beta\epsilon)$
  $/$ $\cosh(\frac{1}{2}\beta\epsilon)$, contributes to the sum in
  Eq.\ (\ref{c2-T}).
  
  The crossing point in the specific heat occurs at the temperature
  $T_+(U)$ for which $C(T,U)$ is independent of $U$. In view of
  Eqs.\ (\ref{txudef}), (\ref{txdef}), and (\ref{c-expansion}), the
  crossing temperature $T_+$ in the limit $U\to0$ is given by the root
  of the equation
  \begin{equation}
    C^{(2)}(T_+)=0,
    \label{cross-cond}
  \end{equation}
  and the specific heat at the crossing point, Eq.\ (\ref{cxdef}), is
  \begin{equation}
    C_+=C^{(0)}(T_+).
    \label{cross-eval}
  \end{equation}
  These equations will be evaluated for several lattices and
  dimensions that are described in the next section.

  \section{Momentum dispersion and density of states}
  \label{nonintsystems}
  
  We consider only one-band systems at half-filling with a symmetric
  density of states on lattices in finite and infinite spatial
  dimensions. For systems with finite bandwidth we set $W/2\equiv1$,
  where $W$ is the bandwidth, while for infinite bandwidth we use a
  unit second moment of the density of states, i.~e.\ 
  $\int\!d\epsilon\,N(\epsilon)\,\epsilon^2\equiv1$.
  
\begin{figure}
    \centerline{\psfig{file=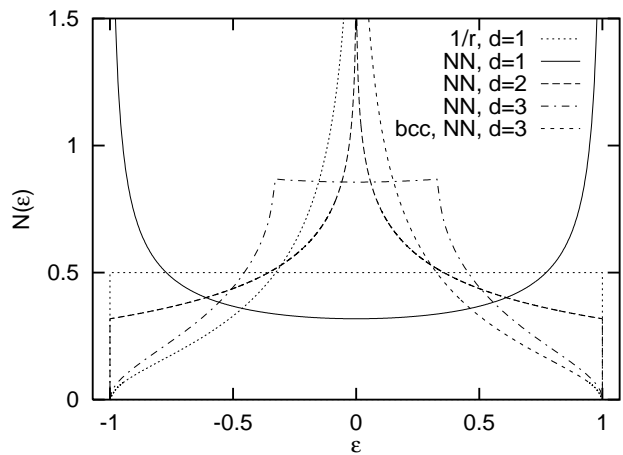,width=1.05\hsize,silent=}}%
    \caption{DOS for several non-in\-ter\-act\-ing systems with 
      next-neighbor and $1/r$-hopping in dimensions 
      $d$ $=$ 1, 2, 3.\label{findospic}}
  \end{figure}

  {\em 1. Finite Dimensions.} For the linear chain, square lattice,
  and simple cubic lattice, i.~e., the hypercubic lattices in
  $d=1,2,3$, we use the tight-binding dispersion $\epsilon_{\bf
    k}=-2t\sum_{i=1}^{d}\cos k_i,$ $|k_i|\leq\pi$, which describes
  nearest neighbor (NN) hopping with amplitude $t\equiv1/2d$.
  Furthermore we study the body-centered cubic (bcc) lattice in $d=3$
  with NN hopping, which can be regarded as a subset of the simple
  cubic lattice with hopping across the space-diagonal, so that
  $\epsilon_{{\bf k}}=-8t\cos(k_x)\cos(k_y)\cos(k_z)$ with
  $t\equiv\frac{1}{8}$.  For these systems we use perturbation theory
  as described in Sec.~\ref{xings}. Finally, for one-di\-men\-sion\-al
  long-range hopping $t(r)\propto1/r$ the known interacting
  dispersion\cite{Gebhard91} can be used instead of perturbation
  theory. The free dispersion is $\epsilon_{k}=tk$, $t\equiv1/\pi$,
  with a constant density of states. Fig.\ \ref{findospic} shows the
  various densities of states used in $d=1,2,3$.

  {\em 2. Infinite dimensions.} We consider first the hypercubic
  lattice and generalized honeycomb lattice with NN hopping. For the
  hypercubic lattice, the hopping must be scaled\cite{Metzner89} as
  $t=1/\sqrt{2d}$ to obtain a nontrivial limit for $d\to\infty$. In
  this case the density of states becomes a Gaussian with unit
  variance, $N(\epsilon)=\exp{(-\epsilon^2/2)}/\sqrt{2\pi}$, whereas
  for the generalized honeycomb lattice\cite{Santoro93} the same
  scaling leads to $N(\epsilon) = |\epsilon| \exp{(-\epsilon^2)}$.  We
  also study the Bethe lattice with infinite connectivity and
  semicircular density of states,
  $N(\epsilon)=\frac{2}{\pi}\sqrt{1-\epsilon^2}$, $|\epsilon|\leq1$.
  
  Furthermore we can take advantage of the fact that in $d=\infty$
  only the density of states of the non-in\-ter\-act\-ing system
  appears in Eq.\ (\ref{e2-infinite-d}). Hence it may be chosen at will
  even when no corresponding dispersion is known.  We consider three
  such functions each containing a tunable real parameter $\alpha>0$.
  This allows us to study the behavior of $C_+$ for a wide range of
  DOS shapes.  The first is the ``metallic'' density of states
  \begin{equation}
    N(\epsilon)=\frac{1+\alpha}{2\alpha}\,(1-|\epsilon|^\alpha),
    \;\;\;|\epsilon|\leq1,
    \label{dosa}
  \end{equation}
  bearing this name due to its finite value at the Fermi energy, while
  the following ``semi-metallic'' density of states
  \begin{equation}
    N(\epsilon)=\frac{1+\alpha}{2}\,|\epsilon|^\alpha,
    \;\;\;|\epsilon|\leq1,
    \label{dosb}
  \end{equation}
  vanishes at the Fermi energy. Furthermore we employ a semi-metallic
  density of states ``with tails'' having infinite bandwidth and unit
  variance
  \begin{equation}
    N(\epsilon)=\frac{c^c}{\Gamma(c)}\,|\epsilon|^\alpha\,
    \exp(-c\,\epsilon^2),\;\;\;c\equiv\frac{1+\alpha}{2}.
    \label{dosc}
  \end{equation}
  It reduces to the DOS for the generalized honeycomb lattice with NN
  hopping in the case of $\alpha=1$.  The important special case of a
  constant rectangular density of states, $N (\epsilon) = \frac{1}{2}$
  for $|\epsilon|\leq1$, is contained in Eq.\ (\ref{dosa}) for
  $\alpha\to\infty$ and in Eq.\ (\ref{dosb}) for $\alpha\to0$.  Note
  that in the limit $\alpha\to\infty$, the DOS in Eq.\ (\ref{dosb})
  approaches two $\delta$-peaks,
  $N(\epsilon)=\frac{1}{2}(\delta(1+\epsilon)+\delta(1-\epsilon))$.
  This particular case is of interest only because in this case the
  integrals in Eqs.\ (\ref{e0}) and (\ref{e2-infinite-d}) can be
  calculated analytically. For general $\alpha$, on the other hand,
  the functions in Eqs.\ (\ref{dosa}) and (\ref{dosb}) model typical
  DOS shapes for finite dimensions.  Several densities of states used
  in $d=\infty$ are depicted in Fig.\ \ref{infdospic}.

  \section{Results for the specific heat at the high-tem\-per\-a\-ture 
    crossing point}
  \label{cxresults}
  
\begin{figure}
    \centerline{\psfig{file=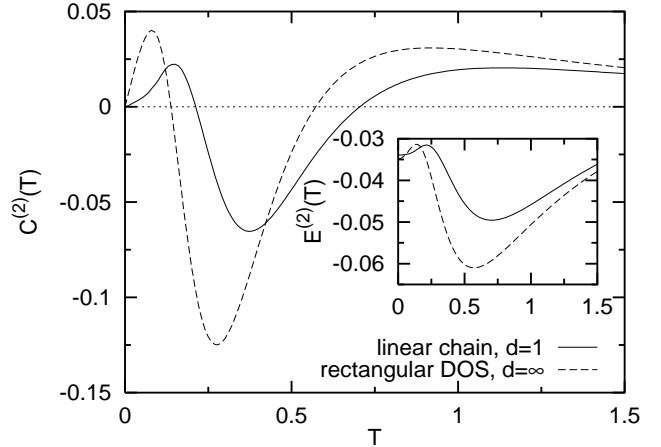,width=1.05\hsize,silent=}}%
    \vspace*{-1mm}
    \caption{Second-order contribution in $U$ to the specific heat, 
      $C^{(2)}(T)$, for the linear chain (with NN hopping) and for the
      constant DOS in infinite dimensions.  The half bandwidth is set
      to 1 in both cases.  The inset shows the correlation
      energy/$U^2$.\label{c2e2pic}}
  \end{figure}

  In this section we present results for the specific heat $C_+$ at
  the high-tem\-per\-a\-ture crossing point for the density of states
  discussed in Sec.~\ref{nonintsystems}. To calculate $C_+$ according
  to Eqs.\ (\ref{cross-cond}) and (\ref{cross-eval}), the integrals
  appearing in Eqs.\ (\ref{c0-T}) and (\ref{c2-T}) have to be
  calculated numerically. We determine them to high precision
  (typically $10^{-8}$) by either Monte-Carlo integration (using the
  VEGAS algorithm\cite{Press92}) or by a high-tem\-per\-a\-ture
  expansion (described in Appendix \ref{e2appendix}). For several
  cases both methods were applied and yielded the same results within
  numerical accuracy.  Typical results for the functions $E^{(2)}(T)$
  and $C^{(2)}(T)$ are shown over a wide temperature range in Fig.\ 
  \ref{c2e2pic} for the linear chain with nearest neighbor hopping and
  for the constant DOS in infinite dimensions.  There is a maximum at
  lower and a minimum at higher temperatures in $E^{(2)}(T)$,
  corresponding to the two zeros of $C^{(2)}(T)$. These are the
  temperatures where the specific heat curves cross for $U\to0$.

\begin{figure}
    \centerline{\psfig{file=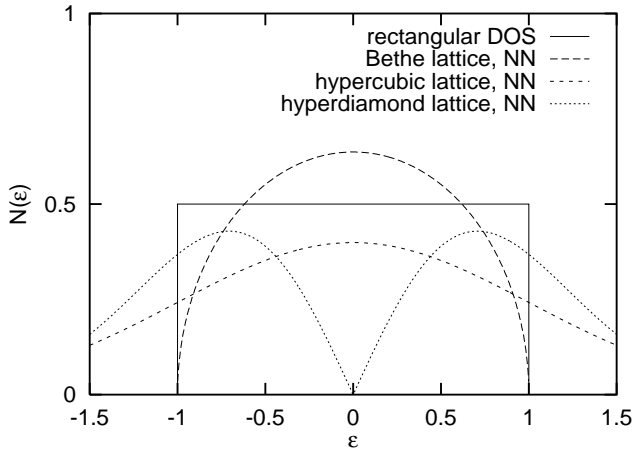,width=1.05\hsize,silent=}}%
    \caption{DOS for several non-in\-ter\-act\-ing systems in 
      infinite dimensions.\label{infdospic}}
  \end{figure}

\begin{figure}
    \centerline{\psfig{file=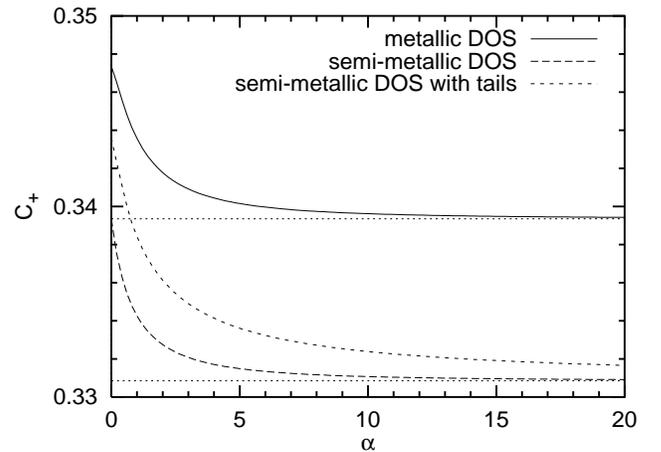,width=1.05\hsize,silent=}}%
    \vspace*{-1mm}
    \caption{$C_+$ for several DOS with a tunable parameter
      $\alpha$, Eqs.\ (\ref{dosa})-(\ref{dosc}), plotted vs.\ $\alpha$.
      The upper and lower dotted horizontal lines represent $C_+$ for
      the constant DOS and the two $\delta$-peak DOS,
      respectively.\label{ctunepic}}
  \end{figure}

  Numerical values for $T_+$ and $C_+$ are listed in
  Table~\ref{cxtxtable}. In view of the drastically different DOS
  shapes (see Fig.\ \ref{findospic} and \ref{infdospic}) it is quite
  remarkable that the values of $C_+$ for these systems are very
  similar, ranging from 0.331 to 0.358.  For the DOS with tunable
  parameter $C_+$ is plotted vs.\ $\alpha$ in Fig.\ \ref{ctunepic}.
  Again, $C_+$ varies only by a small amount although the shapes of
  the DOS change strongly. In particular, for the semi-metallic DOS
  [Eq.\ (\ref{dosb})] $C_+$ lies between the values for the constant and
  two $\delta$-peak DOS, since these are the limits of
  Eq.\ (\ref{dosb}) for $\alpha\to0,\infty$.

  \vspace*{-6mm}
  \begin{table}
    \begin{center}
      \begin{tabular}{l|c|c|c|c}
        system & hopping & $d$ & $C_+$ & $T_+$\\
        \hline
        linear chain&$1/r$&1
        &0.346994&0.561816\\
        linear chain&NN&1
        &0.355547&0.705047\\
        square lattice&NN&2
        &0.352682&0.443585\\
        simple cubic&NN&3
        &0.348327&0.358091\\
        body-cent. cubic&NN&3
        &0.357578&0.241221\\
        hypercubic&NN&$\infty$
        &0.343630&0.847667\\
        hyperdiamond&NN&$\infty$
        &0.338411&0.983569\\
        Bethe lattice&NN&$\infty$
        &0.340906&0.480185\\
        rectangular DOS&undeterm.&$\infty$
        &0.339352&0.571895\\
        two $\delta$-peaks DOS&undeterm.&$\infty$
        &0.330857&1.115358
      \end{tabular}%
    \end{center}
    \vspace*{-6mm}
    \caption{Values of the crossing temperature $T_+$ and
      specific heat $C_+$ at the high-tem\-per\-a\-ture crossing point
      for several $d$-di\-men\-sion\-al Hubbard models. 
      The temperature is given in units of the half bandwidth, 
      except for the hypercubic and hyperdiamond
      lattice, for which the variance of the DOS is set to 
      unity.\label{cxtxtable}}
  \end{table}
  \vspace*{-6mm}

  These results raise the following questions which will be addressed
  in the next section: (i) Why is $C_+$ at the high-tem\-per\-a\-ture
  crossing point so insensitive against changes of the DOS of the
  non-in\-ter\-act\-ing electrons and the spatial dimension? (ii) What
  determines the (small) spread in the values of $C_+$? (iii) Why is
  the value of $C_+$ at the low-tem\-per\-a\-ture crossing point
  varying so much stronger?

  \section{Expansions of $C_+$}
  \label{theory}
  
  We will see that the influence of the DOS and of the crystal lattice
  on $C_+$ can be understood by expanding around the limit of
  $d=\infty$. In addition, in $d=\infty$ the dispersion $\epsilon_{\bf
    k}$ enters only via the DOS $N(\epsilon)$, so that the effect of
  its form on $C_+$ can be studied by expanding in terms of the
  difference between $N(\epsilon)$ and a reference DOS
  $\bar{N}(\epsilon)$.

  \subsection{Influence of the DOS in $d$ $=$ $\infty$}
  \label{deltan}
  
  First we consider infinite-di\-men\-sion\-al systems with an
  arbitrary symmetric DOS $N(\epsilon)$ with finite bandwidth. The DOS
  is compared to a rectangular-shaped DOS with the same bandwidth,
  $\bar{N}(\epsilon)=\frac{1}{2}$ for $|\epsilon|<1$. Their difference
  is characterized by the quantity
  \begin{equation}
    \delta
    N=\int\limits_{-1}^{1}{\!\!}d\epsilon\,
    |N(\epsilon)-\bar{N}(\epsilon)|,
    \label{deltandef}
  \end{equation}
  which will serve as an expansion parameter.  We will expand $C_+$ to
  lowest order in $\delta N$, making use of the known functions
  $\bar{f}_0(x,\beta)$, $\bar{C}^{(2)}(\beta)$, $\bar{C}^{(0)}(\beta)$
  pertaining to $\bar{N}(\epsilon)$, as well as the known crossing
  points for the rectangular DOS,
  $\bar{T}_{+}=1/\bar{\beta}_{+}=0.13801$ and $\bar{C}_{+}=0.44046$ at
  low temperatures, while $\bar{T}_{+}=1/\bar{\beta}_{+}=0.57190$ and
  $\bar{C}_{+}=0.33935$ at the high-tem\-per\-a\-ture crossing point.

  We begin by expanding $C^{(2)}(\beta)$ for small $\delta N$:
  \end{multicols}
  \vspace*{-3mm}\noindent\rule[2mm]{86.36mm}{.1mm}\rule[2mm]{.1mm}{2mm}\vspace*{-2mm}
  \begin{equation}
      C^{(2)}(\beta)=\bar{C}^{(2)}(\beta)
    +4\left[
      \frac{\beta^2}{32}\frac{\partial^2}{\partial\beta^2}\;\beta^2
      \int\limits_0^{\mbox{\ }1}{\!\!}dx\,[f_0(x,\beta)-\bar{f}_0(x,\beta)]
      \bar{f}_0(x,\beta)^3\right]
    +O((\delta N)^2).
    \label{deltancond}
  \end{equation}
  Using standard inequalities it can be shown that the neglected terms
  indeed vanish like $(\delta N)^2$ (for fixed $\beta<\infty$). Now
  $\bar{C}^{(2)}(\beta)$ is expanded around the known crossing point
  $\bar{\beta}_+$ where $\bar{C}^{(2)}(\bar{\beta}_+)=0$. Then the new
  crossing point temperature is determined from the condition
  \begin{equation}
    0\stackrel{!}{=}
    C^{(2)}(\beta_+)
    =
    (\beta_+-\bar{\beta}_+)\bar{C}^{(2)\prime}(\bar{\beta}_+)
    +4\left[
      \frac{\beta^2}{32}\frac{\partial^2}{\partial\beta^2}\;\beta^2
      \int\limits_0^{\mbox{\ }1}{\!\!}dx\,f_0(x,\beta)
      \,\bar{f}_0(x,\beta)^3\right]_{\beta=\bar{\beta}_+}
    +O((\delta N)^2).
  \end{equation}
  To lowest order the shift in $\beta_+$ is hence given by
  \begin{equation}
    \delta\beta_+=\beta_+-\bar{\beta}_+=
    \int\limits_{-1}^1{\!\!}d\epsilon\,N(\epsilon)
    \;\Delta\beta_+(\epsilon) +O((\delta N)^2),
  \end{equation}
  where
  \begin{equation}
      \Delta\beta_+(\epsilon)
      =
      -\frac{\bar{\beta}_+^2}{8\,\bar{C}^{(2)\prime}(\bar{\beta}_+)}
    \left[
      \frac{\partial^2}{\partial\beta^2}\;
      \frac{\beta^2}{\cosh(\frac{1}{2}\beta\epsilon)}
      \int\limits_0^{\mbox{\ }1}{\!\!}dx\,
      \bar{f}_0(x,\beta)^3
      \,\cosh({\textstyle\frac{1}{2}}
      x\beta\epsilon)\right]_{\beta=\bar{\beta}_+}
    \!\!.
  \end{equation}
  \noindent\hspace{22pc}\vspace*{-2mm}\rule[0mm]{.1mm}{2mm}\rule[2mm]{86.36mm}{.1mm}
  \begin{multicols}{2}
  Finally we expand $C^{(0)}(\beta)$ in $\beta-\bar{\beta}_+$ to first
  order and evaluate it for $\beta=\bar{\beta}_++\delta\beta_+$. Thus
  the specific heat at the crossing point $C_+$ is obtained to lowest
  order in $\delta N$ as
  \begin{equation}
    C_+=
    \bar{C}_+
    +\int\limits_{-1}^1{\!\!}d\epsilon\, N(\epsilon)
    \,\Delta C_+(\epsilon)+O((\delta N)^2),
    \label{deltancx}
  \end{equation}
  where
  \begin{eqnarray}
    \Delta C_+(\epsilon)
    &=&
    \frac{\bar{\beta}_+^2\epsilon^2
      }{
      2\,\cosh^2(\frac{1}{2}\bar{\beta}_+\epsilon)}
    \nonumber\\&&\:\:\:\:
    +\,
    \bar{C}^{(0)\prime}(\bar{\beta}_+)\,\Delta\beta_+(\epsilon)
    -
    \bar{C}^{(0)}(\bar{\beta}_+).
  \end{eqnarray}
  The last two equations show how the first order effect on $C_+$ of a
  deviation of $N(\epsilon)$ from a rectangular shape can be
  determined by a single integration.
 
\begin{figure}
    \centerline{\psfig{file=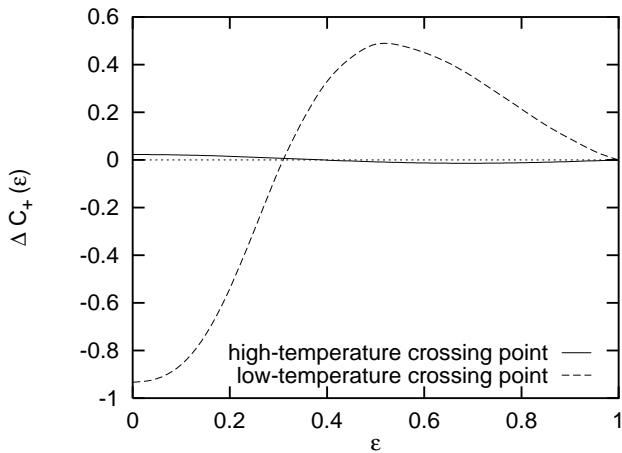,width=1.05\hsize,silent=}}%
    \caption{Weight function $\Delta C_+(\epsilon)$ 
      determining the shift in $C_+$ at the low- and
      high-tem\-per\-a\-ture crossing points for a Hubbard model in
      $d=\infty$ with finite bandwidth, according to
      Eq.\ (\ref{deltancx}). The half bandwidth is set to
      unity.\label{cshift}}
  \end{figure}

  The numerical evaluations of the function $\Delta C_+(\epsilon)$ are
  plotted in Fig.\ \ref{cshift} for both crossing points. The
  amplitudes of the function $\Delta C_+(\epsilon)$ corresponding to
  the low- and high-tem\-per\-a\-ture crossing points are seen to
  differ greatly, i.~e.\ by a factor of about 40. This implies a much
  greater sensitivity towards changes of the DOS and the dimension of
  the value of $C_+$ at the {\em low}-tem\-per\-a\-ture crossing
  point.
  
  It is also clear that $C_+$ is not entirely universal at the
  high-tem\-per\-a\-ture crossing point, since for a general density
  of states $N(\epsilon)$ the integral in Eq.\ (\ref{deltancx}) does
  not vanish. We can in fact estimate the maximum value of the shift
  $\delta C_+=C_+-\bar{C}_+$ for arbitrary $N(\epsilon)$ with finite
  bandwidth, using simple integral inequalities:
  \begin{eqnarray}
    |\delta C_+|&\leq&
    \min\Big(
    a_1,\;
    a_2\cdot
    \max_{0\leq\epsilon\leq1}\!\big[N(\epsilon)\big]
    \Big)
    +O((\delta N)^2),
  \end{eqnarray}
  with $a_1\equiv\max_{0\leq\epsilon\leq1}|\Delta C_+(\epsilon)|$,
  $a_2\equiv\int_{-1}^1{\!}d\epsilon \;|\Delta C_+(\epsilon)|$.  At
  the high-tem\-per\-a\-ture crossing point we find $a_1=0.02268$ and
  $a_2=0.02200$, i.~e.\ to order $O(\delta N)$ we have $|\delta
  C_+|\ll\bar{C}_+$, which is the reason for the insensitivity of
  $C_+$ to changes in $N(\epsilon)$. Furthermore the predicted range
  of values $0.339\pm0.023$ corresponds well to the observed range of
  $0.331\,$-$\,0.358$; see Table~\ref{cxtxtable}. On the other hand,
  at the low-tem\-per\-a\-ture crossing point we find $a_1=0.9330$
  and $a_2=0.7727$, so that $|\delta C_+|\approx\bar{C}_+$.  Hence
  $C_+$ is indeed not confined to a small interval in this case.
  
  To check the validity of this expansion we applied
  Eq.\ (\ref{deltancx}) to several infinite-di\-men\-sion\-al systems
  at the high-tem\-per\-a\-ture crossing point. Results for the Bethe
  lattice, as well as for the metallic DOS {[Eq.\ (\ref{dosa})]} and
  the semi-metallic DOS, {[Eq.\ (\ref{dosb})]} at several values of the
  parameter $\alpha$ are given in Table~\ref{deltantable}. As expected
  we find that the lowest-order approximation in Eq.\ (\ref{deltancx})
  describes the behavior of $C_+$ very well if the deviation from the
  rectangular DOS is not too large. The difference between the exact
  value of $C_{+}$ and the approximate value $\bar{C}_{+}+\delta
  C_{+}$ is due to corrections of order $(\delta N)^2$ and hence is
  typically an order of magnitude smaller than the first order
  correction $\delta C_+$.

  Hence we have shown that for a small change $\delta N$ in the
  density of states the variation of $C_+$ at the
  low-tem\-per\-a\-ture crossing point is large, while at the
  high-tem\-per\-a\-ture crossing point it is small and well-described
  by the first order correction in $\delta N$. This gives quantitative
  answers to the questions posed at the end of Sec.~\ref{cxresults}.

  \subsection{Lattice effects in dimension $d$ $=$ 1, 2, 3}
  \label{oneoverd}
  
  To determine the influence of the lattice dimension on the value of
  $C_+$ we study the hypercubic lattice with NN hopping $t$ and the
  scaling\cite{Metzner89} $t=1/\sqrt{2d}$ and expand $C_+$ in $1/d$.
  Since the large variation of $C_+$ at the low-tem\-per\-a\-ture
  crossing point can already be understood from the results of the
  previous subsection, we will perform this calculation only for the
  high-tem\-per\-a\-ture crossing point.
  
  As discussed in Sec.~\ref{xings}, for $d=\infty$ only the local term
  involving $f_0(x,\beta)$ remains in the lattice sum in
  Eq.\ (\ref{c2-T}).  The $1/d$-corrections are given by the
  contribution from the $2d$ nearest neighbors vectors such as ${\bf
    R}_1=(1,0,0,\ldots)$,\cite{Mueller-Hartmann89}
  \begin{eqnarray}
    \lefteqn{ 
      \sum_m f_m(x,\beta)^4=f_0(x,\beta)^4+2d\,f_1(x,\beta)^4+
      \cdots
      }\nonumber\\&&\:\:\:\:
    =f_0(x,\beta)^4+
    \frac{1}{2d}\left[\frac{2}{\beta}\frac{\partial}{\partial x}
      f_0(x,\beta)\right]^4+
    O(d^{-2})
    ,\label{f0-f1}
  \end{eqnarray}
  where a partial derivative w.r.t.\ $x$ was employed in the second
  line to remove the factor $\cos({\bf k}\cdot{\bf R}_1)$ from
  $f_1(x,\beta)$. Hence the corrections to $C_+$ in order $1/d$ can be
  calculated entirely from $f_0(x,\beta)$. This function contains an
  integral over the DOS $N(\epsilon)$ only, which in turn must be
  expanded in $1/d$.\cite{Mueller-Hartmann89} Then the finite-$d$
  correction to $T_+$ and $C_+$ can be calculated by expanding
  Eqs.\ (\ref{cross-cond}) and (\ref{cross-eval}) to first order in
  $1/d$, with the following result for the high-tem\-per\-a\-ture
  crossing point:
  \begin{eqnarray}\label{cxoneoverd}
    C_+&=&0.343630+0.013599\;\frac{1}{d}
    +{{O}}(d^{-2})
    ,
    \\
    T_+&=&
    \left[
      0.847667+0.082650\;\frac{1}{d}
      +{{O}}(d^{-2})\right]\frac{1}{\sqrt{2d}}
    .
  \end{eqnarray}
  Here the temperature scale has been reset to our previous choice of
  unit half bandwidth in finite dimensions (i.~e., the extra factor
  $1/\sqrt{2d}$ must be omitted\ in order to recover $T_+$ in
  $d=\infty$).  Again this expansion compares very well with the
  numerical results; see Table~\ref{oneoverdtable}.  Note that the
  coefficient of $d^{-1}$ in Eq.\ (\ref{cxoneoverd}) is already much
  smaller than $C_{+}$ in $d=\infty$, which is the reason for the
  insensitivity of $C_+$ to lattice effects, such as Brillouin zone
  shape and momentum conservation.  We expect that for longer-range
  hopping the smallness of the deviations in the value of $C_+$ may
  equally be traced to $1/d$-corrections.
  \section{Conclusion}
  \label{conclusion}
  
  For many correlated electronic systems the curves of the specific
  heat vs.\ temperature obtained for different values of a second
  thermodynamic parameter $X$ are known to
  intersect.\cite{Vollhardt97} For the Hubbard model at half-filling
  the specific heat curves for different values of the Hubbard
  interaction $U$ cross twice, the crossing point at
  high-tem\-per\-a\-tures being remarkably sharp up to intermediate
  values of $U$. We observed that $C_+$, the value of the specific
  heat at this crossing point in the weak-coupling limit, is
  practically the same for a several different lattice systems.  We
  analyzed the origin of this conspicuous feature by calculating $C_+$
  in perturbation theory in $U$. We found the values of $C_+$ at the
  high-tem\-per\-a\-ture crossing point to occur in very small
  interval, i.~e.\ $C_+\approx0.34$ is indeed almost independent of
  dimensionality, crystal lattice, and energy dispersion. This is not
  the case for the crossing point at low temperatures, where $C_+$
  varies on a much larger scale.
  
  Qualitatively, the reason for this difference can be traced to the
  relevant energy scales on which $C(T,U)$ varies. At high
  temperatures, the energy scale for $T$ is essentially determined by
  the bandwidth, i.~e.\ by the hopping amplitude $t$ in the dispersion
  $\epsilon_{\bf k}$. At low temperatures, on the other hand, the
  generation of low-energy excitations (which are responsible for the
  strong enhancement of the low-tem\-per\-a\-ture specific heat and
  $\partial C/\partial U>0$) leads to a renormalized energy scale
  $t\to t_{\text{eff}}\ll t$. The first maximum in $C(T,U)$ occurs at
  a temperature that is of the order of $t_{\text{eff}}$ (see
  Fig.~\ref{c2e2pic}). As a consequence, the first sign change in
  $\partial C/\partial U$ is also linked to $t_{\text{eff}}$, so that
  the intersection of $C(T,U)$ and $C(T,0)$ at low temperatures does
  not occur at any predetermined value.  In contrast, the second sign
  change in $\partial C/\partial U$ is determined only by energy
  scales that also appear in the non-in\-ter\-act\-ing system, leading
  to a nearly universal value for the high-tem\-per\-a\-ture
  intersection of $C(T,U)$ and $C(T,0)$.
  
  To gain a more quantitative understanding we identified two small
  parameters which determine the crossing point values $C_+$. The
  starting point for expansions in these small parameters is the limit
  of infinite dimensions ($d=\infty$).  (i) For $d=\infty$ the
  dependence of $C_+$ on the shape of a DOS $N(\epsilon)$ with finite
  bandwidth is well described by the first-order correction in the
  parameter $\delta N=\int\!d\epsilon\,|N(\epsilon)-\frac{1}{2}|$.
  This parameter is a measure of the difference between $N(\epsilon)$
  and a constant rectangular DOS. It turns out that at the
  high-tem\-per\-a\-ture crossing point this correction is small for
  almost all DOS, while it is large at the low-tem\-per\-a\-ture
  crossing point.  (ii) For hypercubic lattices in dimensions $1\le
  d<\infty$ the value of $C_+$ may be obtained by an expansion around
  $d=\infty$ in powers of $1/d$. At the high-tem\-per\-a\-ture
  crossing point the value of $C_+$ of the $d$-di\-men\-sion\-al
  system is already accurately determined by the first-order
  correction in $1/d$ even for $d$ as low as $d=1$, due to the
  smallness of the prefactor of this term.  These expansions show in
  detail why $C_+$ has an almost universal value at the
  high-tem\-per\-a\-ture crossing point.

  \appendix

  \section{Internal energy at weak coupling}
  \label{e2appendix}

  The internal energy per lattice site $E$ is given by\cite{Fetter71}
  \begin{equation}
    E(T,U)
    =
    T\int{\!\!}d{\bf k}
    \sum_{\omega_n,\sigma}
    e^{i\omega_n \eta}\,
    \frac{\epsilon_{\bf k}+\frac{1}{2}\Sigma_{\sigma}({\bf k},i\omega_n)}
    {G_0 (i\omega_n,{\bf k})^{-1}-\Sigma_{\sigma}({\bf k},i\omega_n)}
    ,\label{e-allg}
  \end{equation}
  where $\omega_n$ denotes fermionic Matsubara frequencies,
  $\eta\to0^+$, and $\hbar\equiv1$. The non-in\-ter\-act\-ing Green
  function is $G_0 (i\omega_n,{\bf k})^{-1}=i\omega_n-(\epsilon_{\bf
    k}-\mu)$, and $\Sigma_{\sigma} ({\bf k}, i\omega_n)$ is the
  self-energy for spin $\sigma$. In the paramagnetic phase the sum
  over spins just gives a factor of two, and the spin indices on the
  self-energy can be dropped.
  \vspace*{2mm}
  \begin{table}
    \begin{center}
      \begin{tabular}{l|r|c|c|c|c}
        DOS & $\alpha\phantom{.0}$ & $\delta N$ & 
        $C_{+,\text{approx}}$ &
        $C_{+,\text{exact}}$ &
        diff.\\
        \hline
        rectangular   &              & 0     & 0.339352 & 0.339352 & 0\\
        Bethe lattice &              & 0.231 & 0.341391 & 0.340906 & 0.14\%\\
        metallic      & 4\phantom{.0}& 0.267 & 0.341271 & 0.340444 & 0.24\%\\
        metallic      & 6\phantom{.0}& 0.207 & 0.340482 & 0.339966 & 0.15\%\\
        metallic      & 8\phantom{.0}& 0.169 & 0.340091 & 0.339743 & 0.10\%\\
        metallic      &10\phantom{.0}& 0.143 & 0.339871 & 0.339621 & 0.07\%\\
        metallic      &12\phantom{.0}& 0.124 & 0.339735 & 0.339552 & 0.05\%\\
        semi-metallic & 0.1          & 0.070 & 0.338349 & 0.338381 & 0.02\%\\
        semi-metallic & 0.5          & 0.296 & 0.335393 & 0.335880 & 0.14\%\\
        semi-metallic & 1\phantom{.0}& 0.500 & 0.333279 & 0.334253 & 0.29\%\\
        semi-metallic & 2\phantom{.0}& 0.770 & 0.331629 & 0.332752 & 0.35\%\\
      \end{tabular}
    \end{center}
    \caption{Comparison of the approximate results for $C_+$ at the
      high-temperature crossing point obtained from 
      the expansion in $\delta N$ [Eq.\ (\ref{deltancx})] with the
      exact values for several DOS in $d=\infty$. The
      expansion is controlled by the parameter $\delta N$ which
      measures the ``distance'' of $N(\epsilon)$ to the rectangular
      DOS, see Eq.\ (\ref{deltandef}). ``Metallic'' and ``semi-metallic''
      refer to the DOS of Eqs.\ (\ref{dosa}) and (\ref{dosb}), respectively.
      The last column shows the difference between
      the approximate and the exact results in percent.
      \label{deltantable}} 
  \end{table}
  \vspace*{3mm}
  \begin{table}
    \begin{center}
      \begin{tabular}{l|c|c|c|c}
        lattice & $d$ &
        $C_{+,\text{approx}}$ &
        $C_{+,\text{exact}}$ & diff.\\
        \hline
        hypercubic           & $\infty$ &  0.343630 & 0.343630 & 0\\
        simple cubic         & $3$      &  0.348164 & 0.348327 & 0.05\%\\
        square lattice       & $2$      &  0.350430 & 0.352682 & 0.64\%\\
        linear chain         & $1$      &  0.357229 & 0.355547 & 0.47\%
      \end{tabular}
    \end{center}
    \caption{Comparison of the approximate results for $C_+$ at the
      high-temperature crossing point obtained
      from the expansion in $1/d$ 
      [Eq.\ (\ref{cxoneoverd})] with the exact values for  
      hypercubic lattices in $d$ dimensions.
      The last column shows the difference between
      the approximate and the exact results in percent.\label{oneoverdtable}}
    \vspace*{-8mm}
  \end{table}

  For a symmetric DOS the chemical potential $\mu$ at half filling is
  given by $U/2$ for all temperatures due to particle-hole symmetry.
  It is useful to define shifted functions
  $\hat{G}_0^{-1}={G}_0^{-1}-U/2$ and $\hat{\Sigma}=\Sigma-U/2$, with
  the new chemical potential fixed at 0. Up to second order in $U$
  $\hat{\Sigma}$ is given by only one Feynman diagram where the lines
  now represent $\hat{G_0}$. We write $\hat{\Sigma} ({\bf k},
  i\omega_n) = U^2 \cdot \hat{\sigma}({\bf k}, i\omega_n) +{{O}}
  (U^3)$, with
  \end{multicols}
  \vspace*{-3mm}\noindent\rule[2mm]{86.36mm}{.1mm}\rule[2mm]{.1mm}{2mm}\vspace*{-2mm}
  \begin{eqnarray}
      \hat{\sigma}({\bf k}_1, i\omega_n)
      =\raisebox{-3mm}{%
        \psfig{file=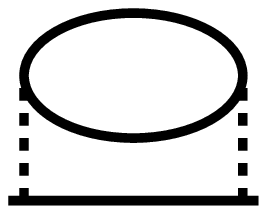,height=8mm,silent=}%
        }
      &=&-T^2 \sum_{\omega_l,\nu_m}
      \int{\!\!}d{{\bf k}_2}
      \int{\!\!}d{{\bf k}_2}
      \int{\!\!}d{{\bf k}_3}\;
    \nonumber\\
    &&
    \times\;
    \hat{G}_0 ({{\bf k}_2}, i\omega_l)
    \hat{G}_0 ({{\bf k}_3},i\omega_l + i\nu_m)\,
    \hat{G}_0 ({{\bf k}_4}, i\omega_n + i\nu_m)
    \sum_{\bf K}
    \delta({\bf k}_1-{\bf k}_2+{\bf k}_3-{\bf k}_4-{\bf K})
    \label{sigma}.
  \end{eqnarray}
  where $\nu_m$ denote bosonic Matsubara frequencies and the sum is
  over reciprocal lattice vectors ${\bf K}$.  Expansion of the
  internal energy [Eq.\ (\ref{e-allg})] in powers of $U$ yields
  Eqs.\ (\ref{e-expansion}), (\ref{e0}), and
  \begin{equation}
    E^{(2)}(T)
    = 
    T\int{\!\!}d{\bf k}\, 
    \sum_{\omega_n}
    \frac{e^{i\omega_n \eta}}{i\omega_n-\epsilon_{\bf k}}
    \left[
      \hat{\sigma}({\bf k},i\omega_n) 
      +2\epsilon_{\bf k}
      \frac{\hat{\sigma}({\bf k},i\omega_n)}
      {i\omega_n-\epsilon_{\bf k}}
    \right]
    .\label{e-prelim}
  \end{equation}
  The frequency summations are carried out as usual,\cite{Fetter71}
  with the result
  \begin{equation}
      T\sum_{n}  
      \frac{
        \hat{\sigma}({\bf k}_1, i\omega_n)
        }{
        i\omega_n-\epsilon_{{\bf k}_1}}
      =-
      \int{\!\!}d{{\bf k}_2}
      \int{\!\!}d{{\bf k}_3}
      \int{\!\!}d{{\bf k}_4}
    \frac{
      \sinh[\frac{1}{2}\beta(
      \epsilon_{{\bf k}_1}-\epsilon_{{\bf k}_2}+ 
      \epsilon_{{\bf k}_3}-\epsilon_{{\bf k}_4})]
      }{
      8[\epsilon_{{\bf k}_1}-\epsilon_{{\bf k}_2}+
      \epsilon_{{\bf k}_3}-\epsilon_{{\bf k}_4}]
      \,\prod_{i=1}^{4}\cosh(\frac{1}{2}\beta 
    \epsilon_{{\bf k}_i})}.
    \sum_{\bf K} 
    \delta({\bf k}_1-{\bf k}_2+{\bf k}_3-{\bf k}_4-{\bf K}),
    \;\label{sigmasum}
  \end{equation}
  Here we eliminate the energy denominator using the identity
  $\sinh(\frac{1}{2}\beta y)=\frac{1}{2}\beta
  \int_0^1{{\!}}dx\,\cosh(\frac{1}{2}x\beta y)$.  Next we rewrite the
  second term in square brackets in Eq.\ (\ref{e-prelim}) as a formal
  derivative $2\epsilon_{{\bf k}_1}\partial/\partial\epsilon_{{\bf
      k}_1}$ of Eq.\ (\ref{sigmasum}).  We make the integrand symmetric
  in all $\epsilon_{{\bf k}_i}$ by shifting ${\bf k}_2$ and ${\bf
    k}_4$ by ${\bf Q}$, where $\epsilon_{{\bf k}+{\bf
      Q}}=-\epsilon_{\bf k}$.  Then the derivative can be replaced by
  $2\beta\partial/\partial\beta$ because only products
  $\beta\epsilon_{{\bf k}_i}$ appear. Taking $\beta$ inside the
  derivative we finally arrive at
  \begin{eqnarray}
      E^{(2)}(T)&=&
      -\frac{{\partial}}{{\partial}\beta}\frac{\beta^2}{32}
      \int\limits_0^{\mbox{\ }1}{\!\!}dx
      \int{\!\!}d{{\bf k}_1} 
      \int{\!\!}d{{\bf k}_2}
      \int{\!\!}d{{\bf k}_3}
      \int{\!\!}d{{\bf k}_4}
    \nonumber\\
    &&
    \times\;
    \frac{\cosh[\frac{1}{2}x\beta
      (\epsilon_{{\bf k}_1}
      +\epsilon_{{\bf k}_2}
      +\epsilon_{{\bf k}_3}
      +\epsilon_{{\bf k}_4})]}{
      \cosh(\frac{1}{2}\beta\epsilon_{{\bf k}_1})
      \cosh(\frac{1}{2}\beta\epsilon_{{\bf k}_2})
      \cosh(\frac{1}{2}\beta\epsilon_{{\bf k}_3})
      \cosh(\frac{1}{2}\beta\epsilon_{{\bf k}_4})}
    \sum_{\bf K} 
    \delta(
    {{\bf k}_1}
    -{{\bf k}_2}
    +{{\bf k}_3}
    -{{\bf k}_4}
    -{\bf K})
    .\label{e2-delta}
  \end{eqnarray}
  In $d=1,2,3$ this simplifies to Eq.\ (\ref{e2-finite-d}), while for
  $d=\infty$ the $\delta$-function can be
  omitted\cite{Mueller-Hartmann89} and the numerator can be replaced
  by $\prod_{i=1}^{4}\cosh(\frac{1}{2}x\beta\epsilon_i)$, leading to
  Eq.\ (\ref{e2-infinite-d}).
  
  The integrals in Eq.\ (\ref{e2-delta}) also factorize if we express
  momentum conservation as a sum over lattice vectors ${\bf R}_m$,
  \begin{eqnarray}
      \sum_{\bf K}
      \delta({\bf k}_1-{\bf k}_2+{\bf k}_3-{\bf k}_4-{\bf K})
    =
    \sum_m
    \exp\left[i
      ({\bf k}_1-{\bf k}_2+{\bf k}_3-{\bf k}_4-{\bf K})
      \cdot
      {\bf R}_m
    \right]
    ,   
  \end{eqnarray}
  leading to Eqs.\ (\ref{c2-T}) and (\ref{fdef}). Since the functions
  $f_m(x,\beta)$ in Eq.\ (\ref{fdef}) cannot be calculated in closed
  form for general $\epsilon_{\bf k}$ we employ a
  high-tem\-per\-a\-ture expansion, which yields
  \begin{eqnarray}
    f_m(x,\beta)&=&\sum_{n=0}^{\infty}\frac{\beta^{n}}{n!}
    \,E_{n}(\textstyle\frac{1+x}{2})\,I_n({\bf R}_m),
    \label{beta-ser}\\
    I_n({\bf R}_m)&=&\int{\!}d{\bf k}\,
    (\epsilon_{{\bf k}})^n
    \exp\left(i\,{\bf k}\cdot{\bf R}_m\right)
    \!,
    \label{rem-int}
  \end{eqnarray}
  where $E_n(x)$ are Euler polynomials. The remaining Brillouin zone
  integrals are calculated as follows. For the $d$-di\-men\-sion\-al
  hypercubic lattice the lattice vectors are ${\bf R}_m=(m_1,\ldots
  m_d)$ with integer $m_i$, so that for NN hopping $t$ we have
  \begin{equation}
    I_n^{\text{hc}}({\bf R}_m)
    =
    (-t)^n\sum_{n_1+\cdots n_d=n}
    {n\choose n_1,\ldots n_d}
    \prod_{i=1}^{d}
    \left[
      \,\int\limits_{-\pi}^{\pi}\!\frac{dk_i}{2\pi}
      \,(2\cos k_i)^{n_i}\,\cos(k_i|m_i|)
    \right]
    .
  \end{equation}
  The integral in square brackets equals ${n_i\choose r_i}$ if $0\leq
  n_i-|m_i|\equiv2r_i$ with integer $r_i$, and zero otherwise.  Hence
  \begin{equation}
    I_n^{\text{hc}}({\bf R}_m)
    =(-t)^n\,n!\;\;\;
    {\sum_{\makebox[0pt]{$\scriptstyle r_1,\ldots,r_d$}}}^{\prime}\;\;\;
    \prod_{i=1}^{d}\frac{1}{r_i!(r_i+|m_i|)!},
  \end{equation}
  where the sum is restricted to $2r_1+\ldots+2r_d=n-M^{\text{hc}}$,
  with $M^{\text{hc}}\equiv\sum_{i=1}^{d}|m_i|$.  Note that
  $I_n^{\text{hc}}$ vanishes if $M^{\text{hc}}>n$ or if
  $n+M^{\text{hc}}$ is odd.  For the bcc lattice in $d=3$ we use the
  hypercubic lattice basis ${\bf R}_m$ and the dispersion given in
  Sec.~\ref{nonintsystems}. We obtain
  \begin{equation}
    I_n^{\text{bcc}}({\bf R}_m)=
    (-t)^n\prod_{i=1}^{3}
    \left[\,\int\limits_{-\pi}^{\pi}\!\frac{dk_i}{2\pi}
      \,(2\cos k_i)^{n}\, \cos(k_i|m_i|) \right]
    =
    (-t)^n\prod_{i=1}^{3}
    {n\choose r_i}
    ,
  \end{equation}
  with $r_i$ defined as above.  In particular $I_n^{\text{bcc}}$
  vanishes if $M^{\text{bcc}}\equiv\mbox{max}_i\,|m_i|>n$ or if
  $n+|m_i|$ is odd.
  
  The advantage of the present high temperature expansion is that
  terms with $M>n$ vanish when expanding to order $\beta^n$, so that
  the lattice sum terminates at $m_i=n$. In addition, lattice
  symmetries can be used to further reduce the computational effort.
  This makes the calculation feasible even in dimension $d=3$ where
  numerical Brillouin zone integration is usually difficult.  At the
  high-tem\-per\-a\-ture crossing point ${{O}}(\beta^{80})$ is
  typically sufficient to obtain $T_+$ and $C_+$ to an accuracy of
  $10^{-8}$.

\end{document}